\documentclass[twocolumn,showpacs,preprintnumbers,amsmath,amssymb,prb]{revtex4}
\usepackage{graphicx, color}
\usepackage{dcolumn}
\usepackage{bm}

\begin{document}

\title{Multilayered cuprate superconductor Ba$_2$Ca$_5$Cu$_6$O$_{12}$(O$_{1-x}$,F$_x$)$_2$ studied by temperature-dependent scanning tunneling microscopy and spectroscopy}

\author{Akira Sugimoto\footnote{asugimoto@hiroshima-u.ac.jp}$^{,1}$}
\author{Toshikazu Ekino$^1$}
\author{Alexander M. Gabovich$^2$}
\author{Ryotaro Sekine$^3$}
\author{Kenji Tanabe$^3$}
\author{Kazuyasu Tokiwa$^3$}

\affiliation{$^1$Graduate School of Integrated Arts and Sciences, Hiroshima University, Higashi-Hiroshima 739-8521, Japan}
\affiliation{$^2$Institute of Physics, National Academy of Sciences of Ukraine, 46 Nauka Ave., Kyiv 03028, Ukraine}
\affiliation{$^3$Department of Applied Electronics, Tokyo University of Science, 6-3-1 Niijuku, Katsushika-ku, Tokyo 125-8585, Japan}

\date{\today}
\begin{abstract}
Scanning tunneling microscopy/spectroscopy (STM/STS) measurements were carried out on a multi-layered cuprate superconductor Ba$_2$Ca$_5$Cu$_6$O$_{12}$(O$_{1-x}$,F$_x$)$_2$.
STM topography revealed random spot structures with the characteristic length $\le 0.5$ nm. 
The conductance spectra $dI/dV(V)$ show the coexistence of smaller gaps $\Delta_S$ and large gaps (pseudogaps) $\Delta_L$. The pseudogap-related features in the superconducting state were traced with the spatial resolution of $\sim$ 0.07 nm. Here, $I$ and $V$ are the tunnel current and bias voltage, respectively.
The temperature, $T$, dependence of $\Delta_S$ follows the reduced Bardeen-Cooper-Schrieffer (BCS) dependence. 
The hallmark ratio 2$\Delta_{S}(T=0)/k_B T_c$ equals to 4.9, which is smaller than those of other cuprate superconductors. Here, $T_c$ is the superconducting critical temperature and $k_B$ is the Boltzmann constant.
The larger gap $\Delta_L$ survives in the normal state and even increases with $T$ above $T_c$.
The $T$ dependences of the spatial distributions for both relevant gaps ($\Delta$ map), as well as for each gap separately ($\Delta_S$ and $\Delta_L$) were obtained.
From the histogram of $\Delta$ map, the averaged gap values were found to be $\bar \Delta_S = \sim 24$ meV and $\bar \Delta_L = \sim 79$ meV.
The smaller gap $\Delta_S$ shows a spatially homogeneous distribution while the larger gap $\Delta_L$ is quite inhomogeneous, indicating that rather homogeneous superconductivity coexists with the patchy distributed pseudogap.
The spatial variation length $\xi_{\Delta_L}$ of $\Delta_L$ correlates with the scale of the topography spot structures, being approximately 0.4 nm. This value is considerably smaller than the coherence length of this class of superconductors, suggesting that $\Delta_L$ is strongly affected by the disorder of the apical O/F.
\end{abstract}

\maketitle

\section{Introduction}

Despite the recent discoveries of new classes of superconductors\cite{klemm12:book,hirsch15:1} with high critical-temperatures ($T_c$'s), including, e.g., magnesium diboride MgB$_2$\cite{nagamatsu}, iron superconductors\cite{kamihara,hosono15}, carbon-based materials\cite{kubozono16}, H$_2$S under high pressures\cite{drozdov15}, etc., the copper-oxide superconductors found in 1986 \cite{bednorz} still remain the main object of the theoretical and experimental research\cite{keimer15:179, kordyuk15:417}.
It can be explained by the highest achieved $T_c$'s at the ambient pressures and various unusual fascinating properties both from the scientific and technological points of view.
However, even those numerous studies could not uncover the mechanism of superconductivity in cuprates as well as the origin of their high $T_c$'s\cite{kresin09:481,kamimura12:677,mukuda12:011008, hashimoto14:483,das14:151}.
Nevertheless, the energy gap structures, especially, the so-called pseudogap (PG) features\cite{hufner08:062501,kordyuk15:417,yoshida16:014513} provide the community with the inestimable information, which helps to examine both the strength of the Cooper-pairing interaction\cite{hashimoto14:483} and the symmetry of the superconducting order parameter\cite{klemm05:801, kirtley11:436}. The quasiparticle energy spectra can also reveal the competing orderings strongly affecting the superconductivity in high-$T_c$ oxides\cite{lee06:17,hashimoto14:483,kordyuk15:417,tajima16:094001}.
The magnitudes of the observed energy gaps in cuprates are of the order of several tens mV, that is very large as compared to their $T_c$'s if the Bardeen-Cooper-Schrieffer (BCS) relationships for the $s$-wave\cite{bardeen57:1175} or $d$-wave\cite{won94:1397} pairings are taken as the reference values.
Furthermore, it has been widely reported that these gap magnitudes were inhomogeneously distributed with the spatial scales of several nm in Bi-based\cite{howald,kmlang,matsuda1,mcelroy,sugimotoPRB2006,machida,fischer07:353,boyer2201,zeljkovic} high-$T_c$ oxides. 
It is remarkable that the gap-value histograms are predominately double-peaked, which seems to be the manifestation of the superconducting gap and pseudogap intertwining\cite{gabovich16:1103}.
Hence, the analysis of the varying inhomogeneities may be useful to understand the cuprate superconductivity because the observed gap spatial distributions include regions of the nanometer scale with gaps being so large, that the corresponding $T_c$'s found on the basis of the BCS relationships substantially exceed the actual ones.
The origin of the inhomogeneities is not known for sure, although recent observations of Bi-based high-$T_c$ superconductors showed that they were closely related to the disorder in the interstitial oxygen and vacancy positions outside the CuO$_2$ conducting layers, the defects acting as the scattering centers\cite{zeljkovic}. One should not also rule out the possibility that the bizarre patterns of the spatial disorder is the consequence of the intrinsic phase separation extensively studied theoretically and most probably revealed experimentally in some of the objects concerned\cite{phillips03:2111,damascelli03:473,lee06:17,fischer07:353,vojta09:699,armitage10:2421}.
However, some problems still remain unclear: (i) whether the inhomogeneous gap distributions are common to all high $T_c$ oxides and (ii) are they driven by the scatter of the Cooper pairing strength or the intensity of the competing interaction leading to the pseudogap formation? The answers to those questions can be obtained by extending the scope of tunnel measurements to other cuprates.

Recently, the high-quality single crystals of the multilayered cuprates with apical fluorine, Ba$_2$Ca$_{n-1}$Cu$_n$O$_{2n}$(O$_{1-x}$, F$_{x}$)$_2$ [F02($n$-1)$n$, $n$ is number of CuO$_2$ planes] have been successfully synthesized \cite{iyo2001,iyo2003,shirage}.
Those compounds seem to be very promising because their $T_c$ exceeds 100 K.
It is also very convenient, that the clean flat surface of BaO/F layer can be obtained by cleaving the single crystal.
Hence, those compounds constitute a class of appropriate materials to investigate their surface nanoscale inhomogeneity depending on their superconducting properties. One can expect that they may serve as another testing ground to manifest gap-spread phenomena revealed earlier in the Bi-based cuprates.
The superconducting properties of Ba$_2$Ca$_{n-1}$Cu$_n$O$_{2n}$(O$_{1-x}$, F$_{x}$)$_2$ can be controlled by changing the number of CuO$_2$ planes $n$, as shown in Fig. 1 (a).
At first, $T_c$ increases with the sheet number $n$, reaches its maximum at $n$=3 and saturates at large $n$\cite{ mukuda12:011008}. It was also shown that the indicated dependence correlates with the charge carrier concentration profiles. (The non-monotonic dependence of $T_c$ on the charge carrier density is inherent, e.g., to the conventional BCS-Eliashberg theory of superconductivity, the decrease at high concentrations being the consequence of the strong electron-phonon-attraction screening in this limit\cite{gabovich78:1115,kresin90:1203}.) Specifically, according to the Nuclear Magnetic Resonance (NMR) experiments, each CuO$_2$ plane demonstrates a different carrier concentrations, i.e., the inner CuO$_2$ planes have lower carrier density than that of outer CuO$_2$ planes. \cite{shimizuNMR_PRL2007,mukuda12:011008}.
The NMR measurements also showed that the inner CuO$_2$ planes tend to become antiferromagnetic while the outer CuO$_2$ ones remain superconducting.
In view of this complexity, it is no wonder that previous point-contact tunnel spectroscopy experiments in F0234 displayed multiple-gap structures, that were attributed to the superconducting gap and pseudogap \cite{miyakawaPCF0234}.
In particular, the pseudogap features were suggested to be induced by the antiferromagnetic state developed in the inner planes below a corresponding critical temperature.
Moreover, the angle resolved photoemission spectroscopy (ARPES) measurements of F0234 revealed double Fermi surfaces \cite{chenARPES}, which the authors considered to originate from the two different types of planes, i.e. inner and outer ones.
Thus, the multiple Fermi surfaces and the multiple gap structures are probably related to the coexisting antiferromagnetism and superconductivity in those multilayer materials.
However, while the differences between the average carrier concentrations in the CuO$_2$ layers were studied in sufficient detail by various means, the existence of intrinsic nanoscale inhomogeneities inside each CuO$_2$ plane is still not understood.

Recently, we carried out Scanning Tunneling Microscopy (STM) experiments on F0256, which identified the apical O and F positions in the BaO/F layers by observing the spot structures \cite{sugimotoPhysC_ISS2008, sugimotoPhysC_ISS2010}.
The inevitable disorder induced by substitutions of the apical O by F (anion replacement) should potentially affects the spatial energy gap distributions.
Therefore, it is natural to expect that the inhomogeneous gap structures in F02($n$-1)$n$ would be observed in the precise STM/Scanning Tunnel Spectroscopy (STS) experiments, and the latter would clarify whether the nanoscale inhomogeneity is the common property of different cuprate superconductors.

In this paper, we present the corresponding systematic observations by the STM/STS technique of temperature ($T$)-dependent gap distributions for the superconductors Ba$_2$Ca$_5$Cu$_6$O$_{12}$ (O$_{1-x}$, F$_x$)$_2$ with six CuO$_2$ layers (F0256, shown in the scheme of Fig. 1(a)).
The results confirmed that the multiple-gap structures with the spatially inhomogeneous magnitude distributions are indeed strongly affected by the disorder at the apical O/F location.

\section{Experimental}
 The single crystals of superconducting Ba$_2$Ca$_5$Cu$_6$O$_{12}$(O$_{1-x}$F$_x$)$_2$ (F0256, $x=$0.79) were fabricated by a high-pressure synthesis technique \cite{iyo2001, iyo2003}.
The typical size of the single crystal is about 0.6 mm $\times$ 0.6 mm $\times$ 0.1 mm. 
The large-size single crystal of this composition is relatively easier to obtain than those of other compositions.
Figures 1(b) and (c) show the examples of the STM image \cite{sugimotoPhysC_ISS2008} and the scanning electron microscope (SEM) image of the single crystal F0256 cleaved surface, respectively.
A rather smooth surface is found by the SEM observation [Fig. 1(c)], which is suitable for STM measurements.
The STM image [Fig. 1(b)] shows atomic arrangements with the spatial period of $\sim$ 0.38 nm (indicated by dashed lines).
These structures should be attributed to the oxygen sites on the CuO$_2$ sheet and Ba/F sites (indicated by circles), the result being basically similar to that obtained earlier for F0245 \cite{sugimotoPhysC_ISS2010}.
We note that such atomic arrangements are observed only inside the area between the large spot structures [shown in Fig. 2 (a)].
The $T_c \simeq$ value of 70 K was determined by measuring the $T$ dependence of the magnetic susceptibility as shown in Fig. 1(d).
The $T$ dependence of resistivity confirms this result.
The x-ray diffraction (XRD) and electron probe micro-analysis (EPMA) measurements confirmed the adopted six-layered CuO$_2$ phase indicated as the schematic model in Fig. 1(a).
The composition ratio determined by EPMA revealed that the average Ba:Ca:Cu:F ratio was 2.00:5.13:5.78:1.59, which is in a good agreement with the ideal stoichiometric ratio for this compound.

The STM equipment used in this experiment is commercially based system (Omicron LT-STM) with upgraded modifications \cite{ekino2007JPCS, sugimotoLBH}.
The single crystals were cleaved at 77 K under the ultra-high vacuum atmosphere of $\sim 10^{-8}$ Pa.
The exposed surface is considered to be the BaO/F layer as shown in Fig. 1(a).
The Pt/Ir tip was cleaned by the high-voltage field emission process with the Au single crystal target just prior to the measurements.
The STM/STS observations were carried out in the temperature range of 4.9 K to 290 K using the heating system in the ultra-high vacuum of $\sim 10^{-8}$ Pa. 
Before STM/STS observations, the local barrier height (LBH; work-function) was measured on the basis of the relationship between the tunneling current ($I$) and the tip-sample distance ($z$) ($I-z$ method \cite{sugimotoLBH}).
Fig. 1(e) shows the simultaneously obtained LBH map of F0256 surface with 5 nm $\times$ 5 nm measurement area.
The histogram of LBH data extracted from the map are presented in Fig. 1(f), demonstrating sufficiently large values (3$\sim 4$ eV) to carry out measurements in the tunneling regime.
$dI/dV$ curves were obtained by numerical differentiation of the measured $I-V$ characteristics with the spatial interval of $\sim$ 0.07 nm. 
The voltage at the sample is considered as a reference one.

\section{Results and discussion}
Figure 2(a) demonstrates an STM image (topography) of the cleaved F0256 single crystal surface at $T=4.9$ K for $V=+0.3$ V and $I=0.25$ nA.
The randomly distributed bright spots (indicated by black dots) of the sub-nanometer size are clearly observed over the whole scanned area.
Such large spot structures have been already reported in our previous communications \cite{sugimotoPhysC_ISS2008,sugimotoPhysC_M2S,sugimotoPhysProc_ISS2013}.
These spots were suggested to correspond to the non-replaced apical oxygen of topmost BaO/F layers. 
This conclusions stems from the observation that the fraction of such spots in the STM images correlates with the O/F ratio obtained by the element analysis. 
Indeed, the spot density obtained by the direct counting was about $\sim $28\% of apical O/F atomic sites, being in the fairly good agreement with the atomic ratio of apical O $(1-x) \simeq 21$ \% found by the EPMA element analysis \cite{sugimotoPhysC_ISS2008}.
It should be noted that clear periodic atomic structures were not visible in this frame.
However, we observed the atomic corrugations with the period of $\sim$ 0.38 nm inside the interval between the spot structures, as is shown in Fig. 1(b). 
Additionally, our previous study of the other compound F0245 showed the existence of apparent atomic-lattice structures with such spots and the fourfold cross shaped clusters of the spots\cite{sugimotoPhysC_ISS2010}.
Empirically from our experiments, such random spot structures were sometimes observed on contaminated oxide surfaces.
However, the LBH measurements, which give a large value (3$\sim 4$ eV) ensuring the ideal vacuum tunneling regime, testify that the observed surface areas were not contaminated. 
The observed random spot structures, which are distributed in the BaO/F layers, cause the disordering outside the ``superconducting" CuO$_2$ planes. 
Nevertheless, they may affect the electron states and, hence, the superconducting gap in the quasiparticle spectrum.
To check the possibility and strength of such disorder effects, we carried out the STS measurements and obtained numerous $dI/dV$ spectra at various locations, the results, as are well known, being proportional to the local density of electron states (LDOS). 


In Fig. 2(b), the $dI/dV$ versus $V$ dependence, averaged over the whole area shown in Fig. 2(a), is represented.
One can clearly recognize two kinds of gap structures in the positive-bias branch: a larger-gap peak (dubbed as $\Delta_L$) and the inner sub-gap kink feature (dubbed as $\Delta_S$), both indicated by arrows in Fig. 2(b).
The larger gap $\Delta_L$ is about $\sim $ 80 meV being defined by the corresponding peak position at $V>0$.
On the other hand, the magnitude of $\Delta_S$, which is defined by the kink feature location for the same positive voltage branch, is about $\sim$ 20 meV. 
The revealed peaks, corresponding to $\Delta_S$, are not as pronounced as those sometimes seen in Bi-based cuprates. 
Nevertheless, the peaks at energies $\Delta_S$ are regularly and reliably observed, being visible for both positive and negative biases. 
The absence of sharp BCS-like peak structures at the edge of the superconducting gap regions are typical for cuprate superconductors, the higher $T_c$ of which does not always correlate with the apparent peaks in the conductance $dI/dV$ spectra.
For example, this is true for YBa$_2$Cu$_3$O$_{7-d}$ (YBCO with high $T_c$ of 93 K), where the $dI/dV$ spectra do not show distinct gap-edge features. In particular, the area-averaged spectra did not show sharp peaks (Ref. [\onlinecite{deroo02:PRL097002}]).
In this connection, it is also worthwhile to consider the ``layer resolved" STS measurements on Bi2212 \cite{lv15:PRL237002}, which demonstrate the multiple gap structures on Bi2212. In particular, the $dI/dV$ curve for ``CuO$_2$ sheet" in Bi2212 turns out to be very similar to our results on F0256.
The asymmetric $dI/dV$ curves for the strongly underdoped ($p\sim$ 6 \%) samples of Bi2212 \cite{hamidian16:Nphys3519} are of the same kind as ours as well. There are two main possible reasons of the observed modest gap-edge manifestations in high-$T_c$ materials. First, is the $d$ symmetry of the superconducting order parameter in cuprates, which is widely accepted as being the case\cite{kirtley11:436} (although the situation is not at all unambiguous\cite{klemm05:801}). Then, the gap anisotropy with nodes leads to the much more smooth behavior of $dI/dV$ both in symmetric and non-symmetric tunnel junctions involving $d$-wave superconductors\cite{won94:1397} than for the isotropic BCS superconductors. The second reason is the intertwining between superconductivity and another long-range order responsible for the pseudogaps. This is exactly what is observed here and in experiments cited above\cite{deroo02:PRL097002,lv15:PRL237002, hamidian16:Nphys3519}.
From these facts, such smeared $\Delta_S$ signals should be regarded to reflect the essential intrinsic features of the sample LDOS rather than the poor sample quality or a contamination influence.

Figs. 2(c) $-$ (e) demonstrate the evolution of the $dI/dV(V)$ profiles along the arrows 1$-$3 in Fig. 2 (a) measured with the increment of $\sim$ 0.07 nm.
Almost all $dI/dV(V)$ curves reveal the asymmetric behavior with the apparent peaks in the positive bias branch in the range $V=+50 \sim +100$ mV, as is shown by the dotted lines in Fig. 2(d).
Such peculiarities are considered to be of the same kind as the average gap $\Delta_L$ clearly seen in Fig. 2(b). 
The asymmetry of the current-voltage characteristics is typical for cuprates \cite{fischer07:353} and may be caused by various reasons \cite{gabovich10:681070}. 
In many curves, the less-pronounced peak or kink structures can be also found for the negative bias branch in the range $V =-50 \sim -100$ mV.
The smaller- (inner-) gap structures are visible simultaneously with the larger-gap features but at substantially lower positive voltages of $V= +10 \sim +30$ mV, as is shown by the dashed marks in Fig. 2(e). 
Those peculiarities are attributed to the gap $\Delta_S$ indicated in Fig. 2(b).
It is remarkable that both coherent-peak positions and heights vary along the surface at small distances in the range of $\sim 1-2$ nm.
The results accumulated in Figs. 2(c)$-$2(e) testify that the double-gap tunnel conductance spectrum is observed at each location point, indicating the coexistence of $\Delta_L$ and $\Delta_S$ at least within the spatial resolution of the measurements ($\leq 0.07$ nm).


To obtain more insight into the origin and character of these multiple-gap spectra, we visualize the two-dimensional (2D) spatial gap distribution ($\Delta$ map) obtained from the $dI/dV $ measurements, like those shown in Figs. 2(c)$-$(e).
In Fig. 3(a), we depicted the $\Delta$ map created at $T=$4.9 K and corresponding to the STM image area of Fig. 2(a).
The $\Delta$ values were defined by the lowest peak locations at the positive bias branch.
If there were two gap peaks for a single $dI/dV(V)$ spectrum, the map point is attributed to the lower one $\Delta_S$ value rather than to the larger one $\Delta_L$. Hence, the $\Delta$ map includes both gap values, namely, $\Delta_L$ and $\Delta_S$.
In Fig. 3(b), such points that correspond to the multiple-gap spectra are indicated by the gray-scaled pixels.
We note that some $dI/dV(V)$ curves exhibit the peculiarity at the smaller gap $\Delta_S$ location as a kink rather than the coherence peak at the gap edge.
Such features were not included into the map of Fig. 3 (b). 
The $\Delta_S$ distribution including the kink structures ($\Delta_{S-kink}$) will be shown and discussed later. (See Fig. 6 (a-1).)
The map sections corresponding to the multiple-gap electron spectrum were much smaller than the single-gap areas.
Fig. 3(c) includes the spatial distribution of $\Delta_L$ ($\Delta_L$ map) defined by the maximum-peak voltage at the positive bias branch, as is shown in Fig. 2(b) by the arrow and notation $\Delta_L$.
The $\Delta_L$ map exhibits the distribution similar to that appropriate to the general $\Delta$ map shown in Fig. 3(a).
In addition, the $\Delta_L$ map reveals patch-like structures with a smooth spatial variation, which are similar to that found for Bi-based and some other families of cuprate superconductors\cite{campi15:359}.
However, in contrast to the latter, one can recognize that the length scale of the spatial variation depicted in Fig. 3 is rather small, about $\le \sim$1 nm.
This length scale is comparable to that of the spot structures, as is seen in the topography of Fig. 2(a).
The character of the gap-distribution length scale will be discussed in more detail below.
The histogram of $\Delta$ map (Fig. 3(a)) is shown in Fig. 3(d).
One can readily see that this histogram has a structure with distinct two-peaks, as is indicated by the two arrows in Fig. 3(d). 
Such a form testifies that the $\Delta_S$ and $\Delta_L$ are entirely separated at the dividing energy of $E\simeq 40$ meV, so that both distributions should be associated with different types of gaps. 
The partitioning into low and high energy gaps of non-similar nature is well-known for cuprates, where high-energy gaps are traditionally called pseudogaps\cite{hufner08:062501,kordyuk15:417,yoshida16:014513} and are most probably the consequence of the charge density wave (CDW) formation, recently observed by various methods\cite{gabovich16:1103,ekino16:445701}.
The $\Delta_L$ distribution is rather wide, in the range from $\Delta_L$ = 40 mV to 140 mV.
The average value of the smaller $\Delta_S$ is $\bar{\Delta_S}\simeq$ 24 meV with the standard deviation $\sigma \simeq$ 5 meV, while that of the larger one is $\bar{\Delta_L} \simeq$ 79 meV with $\sigma \simeq$ 17 meV.
Figures 3(a) and 3(b) demonstrate that the $\Delta_S$-covered area in the map consists of disconnected fragments covering only about $\sim 7.5$ \% of the whole surface.
This small area is shown in Fig. 3(c) as dark spots distributed inside the major $\Delta_L$-covered area.


The temperature-dependent STS measurement results are presented in Fig. 4. 
They include the typical $dI/dV$ line spectra, $\Delta$ maps, and the $\Delta$-map histograms, depicted in upper, middle, and bottom panels, respectively. 
The selected $T$ values are (a) 4.9, (b) 15, (c) 45, (d) 60, and (e) 77 K, thus including temperatures both below and above $T_c$.
The relevant notations are the same as those of Fig. 3 (a).
The spectra were measured for 64 $\times$ 64 points at $T=$4.9 and 15 K, while at $T=$ 45, 60, and 77 K the number of points was reduced to 32 $\times$ 32 in order to maintain the stable measurement conditions.
The results obtained at $T=$4.9 K and represented in Fig. 4(a) are the same as in Fig. 3 and are included once more for comparison.
The locations of the $dI/dV$ profile sets for each upper panel are indicated by black arrows in the corresponding $\Delta$ maps (middle panels).
The $dI/dV$ line profiles of Fig. 4(b) ($T=15$ K) demonstrate the clear-cut peak structures at $\Delta_L$ as recognizable as those for $T=$4.9 K.
Similarly, the histogram at $T=$15 K reveals the double-peak gap distributions (lower $\Delta_S$ and higher $\Delta_L $) with the border energy of $ \sim 40$ meV, which are as clearly observed as same as those at $T=$4.9 K.
The average values of the relevant gaps are $\bar{\Delta_S}$($T=$ 15 K) $\simeq$ 20 meV and $\bar{\Delta_L}$($T=$ 15 K) $\simeq$ 62 meV.
On the other hand, the $dI/dV$ line spectra at $T=$ 45 K and 60 K are thermally smeared but the gap peaks still can be recognized at about $V \simeq \pm$ 50 mV.
The histogram at $T=$ 60 K also exhibits the double-peak distribution, with the energy of $\sim 30$ meV being an approximate point dividing $\Delta_S$ and $\Delta_L $.
If we accept the border between $\Delta_S$ and $\Delta_L $ distributions as $\Delta $=30 meV, the average values of the gaps involved become $\bar{\Delta_S}$ (45 K) $\simeq$ 20 meV, $\bar{\Delta_S}$(60 K) $\simeq$ 17 meV, $\bar{\Delta_L}$(45 K) $\simeq$ 63 meV, and $\bar{\Delta_L}$(60 K) $\simeq$ 65 meV.
In Fig. 4 (e), the STS results are shown for $T=$ 77 K, i.e. above $T_c$.
The corresponding $dI/dV$ spectra reveal large shallow depressions with remnants of broad gap peaks around $V \sim$ 100 mV.
From the histogram at $T=$ 77 K, the average gap can be estimated as $\bar{\Delta}$(77 K) $\simeq$ 104 meV.
At the same time, the histogram at $T=$ 77 K shows no traces of the double-peak patterns. 
Since $\Delta_S$ vanishes above $T_c$, it is reasonable to suggest that it is the true superconducting gap, whereas the interpretation of the other gap $\Delta_L $, which survives in the normal state, needs further discussion.


In order to accurately obtain the $T$ dependence of $\Delta$, it is necessary to measure the temperature evolution of $dI/dV(V)$ curves both below and above the critical temperature, smoothly crossing this crucial point.
The results of such measurements carried out in the bias voltage range of $|V| <$ 200 mV and the $T$ range of $T=$ 4.9 K to 284 K are depicted in Figs. 5 (a) and (b).
In Fig. 5(b), the presented $dI/dV(V)$ curves were derived by averaging the whole measured $dI/dV$ set of Figs. 4(a) to 4(d). 
Each $dI/dV(V)$ curve shown in Figs. 5 (a) and (b) was obtained by dividing the measured $dI/dV$ conductance by $V=$120 mV ($G(V)/G(120$mV)=1 at $V=120$ mV). For clarity of the comparison, the curves were offset. 
Looking at Fig. 5(a), one can clearly recognize that the LDOS depletion associated with the large $\Delta_L$ indeed survives far above $T_c$ up to about $T \sim$ 200 K.
The persistence of the gap-like electron spectrum behavior above $T_c$ is consistent with the indicated above pseudogaps, which clearly manifest themselves in Bi-based cuprate superconductors \cite{keimer15:179,yazdani,ekinoBi,matsuda2, kordyuk15:417,kurosawa10:094519}.

The $T$ evolution in the $T$ range 4.9 K to 70 K of averaged $dI/dV(V)$ spectra is shown in Fig. 5(c) for low bias voltages $|V|<$40 mV, including the $V$ range where the smaller (superconducting) gap $\Delta_S$ clearly manifests itself. 
This data set makes it possible to find the $T$ dependence of $\Delta_S$\cite{sugimotoPhysC_ISS2008}.
In order to make results quantitative, our experimental $dI/dV(V)$ dependencies were fitted by the theoretical formula for the tunnel conductance.
Namely, the fitting curves were based on the basic Dynes phenomenological function\cite{dynes78:1509}. 
However, the latter was modified by the weighting function reflecting the adopted $d_{x^2-y^2}$-wave superconducting order parameter symmetry\cite{won94:1397} and with an additional function making allowance for the linear background function manifesting itself for cuprates\cite{kirtley92:336}, 
\begin{equation}
\frac{dI}{dV}\propto \int^{\pi/2}_{-\pi/2} \left|{\rm Re}\left[\frac{eV- i\Gamma}{\sqrt{(eV-i\Gamma)^2-\Delta_{S fit} ^2 \cos ^2 2\theta}} \right] \right| d\theta + kV,
\end{equation}
where $\Gamma$ and $k$ are the Dynes broadening parameter and a constant, respectively\cite{sugimotoPhysC_ISS2008}.
The variable $\theta$ is the angle measured from the anti-nodal line on the 2-dimensional CuO$_2$ plane.
Thin solid lines in Fig. 5(c) correspond to the fitting curves at each temperature.

At 4.9 K, the fitting results are as follows: $\Delta_{S fit}$ = 14.7 meV, $\Gamma$ = 4.4 meV and $k=5.15 \times 10^{-3}$ (V$^{-1}$).
As $T$ increases, the calculated superconducting gap $\Delta_{S fit}$ remains almost constant up to $\sim$ 40 K, for example, $\Delta_{S fit}$(10 K) = 15 meV and $\Delta_{S fit}$(40 K) = 16 meV.
However, above $T$ = $\sim $ 50 K, $\Delta_S$-related peculiarities in the $dI/dV$ curves are quickly smeared and the value of $\Delta_{S fit}$ begins to gradually decrease. 
In Fig. 5(d), the $\Delta_{S fit}$ values found at various temperatures are plotted by solid diamonds.
The dashed line in Fig. 5(d) represents the $T$ dependence of the gap given by the conventionally reduced BCS theory, which (in this coordinates) is quite similar for $d_{x^2-y^2}$-wave and $s$-wave symmetries of the order parameter\cite{voitenko10:20}.
One sees that the curve $\Delta_{S fit}(T)$ follows the $d$-wave weak-coupling BCS law of the corresponding states\cite{won94:1397} (the dashed curve).
The magnitudes of the $\bar{\Delta_S}$ and $\bar{\Delta_L}$ are plotted together in Fig. 5(d) as the open diamonds and the open triangles, respectively, showing that $\bar{\Delta_S}$ slightly decreases with growing $T$, while $\bar{\Delta_L}$ exhibits no conspicuous changes in the range 4.9 K $\leq T \leq$ 60 K.
All the fitting values of $\Delta_{S fit}$ are smaller than the averaged ones $\bar{\Delta_{S}}$. 
At the same time, the empirical relationship $\Delta_{S fit}+ \Gamma = \bar{\Delta_{S}}$ \cite{Ekino_Bi_PhysRevB1989} is approximately valid for the results obtained. 
The role of broadening factors was also discussed in more detail for tunnel measurements of superconducting $\alpha$-K$_x$TiNCl and $\beta$-HfNCl$_y$\cite{sugimotoPRBMNCl}.
The huge discrepancy between $\Delta_{S fit}$ and $\bar{\Delta_{S}}$ cannot be understood on the basis of the dissipation-related broadening $\Gamma$ and its thermal counterpart. 
Thus, its nature remains obscure. 
That is why we considered reasonable to analyze $\Delta_{S fit}$ rather than $\bar{\Delta_{S}}$ to elucidate the actual temperature dependence of the gap $\Delta (T)$, whatever broadening and disorder factors influence the gap-value distributions.
Hence, a true superconducting energy gap identified with the calculated $\Delta_{S fit}$ takes the value $\sim$ 15 meV at $T=0$. 
This leads to the gap $vs.$ $T_c$ ratio 2$\Delta_{S}(0) /k_B T_c \sim $ 4.9, which apparently exceeds the weak-coupling values of this important mean-field parameter both for $s$ and $d$-wave superconductors. 
It means that the strong-coupling effects are important although not crucial in our case, so that the use of the phenomenological treatment\cite{dynes78:1509} is justified. Here, $k_B$ is the Boltzmann constant.
The value of this benchmark ratio $\sim $ 4.9 is smaller than that of Bi$_2$Sr$_2$CaCu$_2$O$_{8+\delta}$ (Bi2212, = 7 $\sim$ 10) or La$_{2-x}$Sr$_x$CuO$_4$ (= 5 $\sim$ 10), but is comparable to that of YBa$_2$Cu$_3$O$_y$ (= 4 $\sim$ 6) and electron-doped Ln$_{2-x}$Ce$_x$CuO$_4$ (= 4 $\sim$ 7, Ln= lanthanides) \cite{sugimotoPRB2006,ekinoNCCO,zimmers,kato,maki}.
The unusually small ratio as compared to that of the similar Bi2212 may be associated with difference in inhomogeneous $\Delta_L$ and $\Delta_S$ distributions or to the change of the ratio 2$\Delta_{S}(0) /k_B T_c $ due to the influence of the pseudogap (CDW gap) $\Delta_L$\cite{gabovich10:681070}.

The $T$ dependences of the larger gap $\Delta_L $ are depicted in the upper part of Fig. 5(d). 
There are two kinds of the measured quantities, which may represent the actual gap. 
The first one is $\bar{\Delta_{L}}$, being the averaged value from the histogram describing the $\Delta_{L}$ group of Figs. 4 (red-open triangles), and the other one (red-solid triangles) is directly obtained as the outer peak voltage at the positive bias of the averaged $dI/dV(V)$ curve, such as shown, e.g., in Fig. 5 (a).
It comes about that the gap $\Delta_L$ persists both below and above $T_c \sim$ 70 K and its magnitude increases from 60 to $\sim$ 110 meV with growing $T$. 
In analogy with Bi-based cuprates, the quantity $\Delta_L$ is suspected to be a pseudogap (CDW gap) competing with superconductivity\cite{gabovich10:681070,kordyuk15:417,gabovich16:1103,voitenko10:20}. 
Hence, the initial increase of $\Delta_L$ when superconductivity is suppressed near $T_c$ is quite natural. 
On the other hand, the further continuous enhancement of $\Delta_L$ at higher temperature is counterintuitive, since in the mean-field scenario the CDW order parameter behaves similarly\cite{gruner94:book} to its superconducting analog, which was clearly observed, e.g, for dichalcogenides\cite{shen07:216404} and trichalcogenides\cite{sinchenko99:4624,ekino1986}. 
It should be noted that the growth of the PG with $T$ far above $T_c$ was also observed for Bi(Pb)$_2$Sr$_2$Ca(Tb)Cu$_2$O$_{8+\delta}$\cite{kordyuk09:020504}. 
However, in the latter material the enhancement gave way to the decrease at ever larger temperatures, making nonmonotonic the overall PG behavior. 
Peculiarities intrinsic to the presented here larger gap dependence as well as data of Ref. [\onlinecite{kordyuk09:020504}] might be caused either by the unusually strong dielectric order parameter fluctuations in those quasi-two-dimensional oxides \cite{kivelson03:1201,fradkin15:457} or the interplay between the pseudogap and CDW gaps being different phenomena in this case, coexisting in the non-superconducting region of the phase diagram\cite{keimer15:179,kordyuk15:417}. 
This speculation is supported by the recent observation of the CDW-gap nonmonotonic behavior (analogous to that found in high-$T_c$ oxides\cite{kordyuk09:020504}) in the superconducting layered dichalcogenide $2H$-NbSe$_2$ as well\cite{borisenko09:166402}.
The very interpretation of the gap $\Delta_L $ in F0256 as the CDW one is tentative and is based on the analogy between this superconducting cuprate and its analogues, where CDWs were first guessed to exist and later on discovered directly by x-ray scattering. For instance, CDWs were discovered in YBa$_2$Cu$_3$O$_{7-\delta}$\cite{achkar12:167001,forgan15:10064}. By CDWs in cuprates we mean not only charge-carrier static waves but also the concomitant periodic lattice distortions minimizing together the free energy of the whole metallic system \cite{gruner94:book, rossnagel11:213001}.

In Fig. 5(d), one can easily see the difference in energy between $\Delta_L$ and $\Delta_S$, specifically, $\Delta_L$ is about 4 to 5 times larger than $\Delta_S$.
However, in contrast to $\Delta_S$, which in the usual way disappears below the critical temperature of the superconducting transition $T_c \sim 70$ K, the competing gap $\Delta_L$ survives in the normal state, up to $\sim$ 200 K.
We note, that this is in line with the pseudogap (the CDW gap, in our interpretation\cite{gabovich10:681070,gabovich16:1103,voitenko10:20}) behavior in a number of cuprate superconductors, especially in Bi-based ones, when the larger gap manifests itself above $T_c$, as was confirmed by various experimental methods \cite{keimer15:179,yazdani,ekinoBi,matsuda2,hufner08:062501,kordyuk15:417,yoshida16:014513,mesaros16:1266}.
In particular, in our previous break-junction (BJ) tunneling experiment, the temperature dependence of the normal-state gap magnitude varied up and down in the $T_c$ neighborhood and persisted up to almost the room temperature \cite{ekinoBi}.


As was mentioned above, the magnitudes of both relevant gaps in high-$T_c$ oxides demonstrate substantial spatial scatter\cite{howald,kmlang,matsuda1,mcelroy,sugimotoPRB2006,machida,fischer07:353,boyer2201,zeljkovic,miyakawa05:225,gabovich16:1103,ekino16:445701}. 
In our samples, the length scale of the $\Delta$ variation, especially that of the $\Delta_L$ is very small.
In order to clarify the correlation between the spot structures in the topography and the spatial $\Delta$ variations, the continuous spatial variations of $\Delta_S$ also should be explored.
However, the $\Delta_S$ peak is so small that the distribution of $\Delta_S$ is discrete.
Hence, we additionally-estimate the distribution of $\Delta_S$ including the ``kink" structure $\Delta_{S-kink}$, which was not taken into account in the $\Delta_S$ map of Fig. 3(b). 
The definition of the $\Delta_{S-kink}$ value is the inflection point of the kink structure around $V$= +5$\sim$40 meV.
Fig. 6 (a-1) displays the $\Delta_{S-kink}$ map on the squared area of the topography of Fig. 2 (a) and is shown together with with the contour plots of the topography.
The $\Delta_L$ map with the contour plots of the topography is also depicted in Fig. 6(a-2), reproducing the bottom part of Fig. 3 (c).
The line-cut profiles of $\Delta_{S-kink}$ and $\Delta_L$ taken at the same positions along the arrows shown in Figs. 6(a-1) and (a-2) are plotted in Fig. 6(b) along with the topography profiles.
The $\Delta_{S-kink}$ was seen uniformly in the almost whole area except several positions of the spot structure.
These results indicate that the quantity $\Delta_{S-kink}$ truly characterizes the superconducting states and is homogeneously distributed, while the larger gap $\Delta_L$ is quite inhomogeneous and seems to be affected by the spot structures.
As stems from the profile plots, the $\Delta_L$ values do not exactly correlate with the topography height. 
For instance, the spot position A is in the small-$\Delta_L$ region, while the spot B is found in the large-$\Delta_L$ area.
However, the steep spatial variation of $\Delta_L $ does occur in the neighborhood of the spot structures.
Thus, the gap $\Delta_L$ is strongly influenced by spot structures contrary to its counterpart $\Delta_{S-kink}$.
Hence, the disorder appropriate to the BaO/F layers strongly modifies the pseudogap, while the superconducting gap of $\Delta_S$ is not conspicuously influenced. It means that the sample inhomogeneity is linked to the non-homogeneous spatial distribution of the pseudogaps, the latter proven previously to be intrinsic to Bi-based ceramics\cite{boyer2201}.
To treat the spatial inhomogeneity of $\Delta_L$ quantitatively, we estimated the full width at the half-maximum height (FWHM) of the $\Delta_L$ map autocorrelation, $\xi_{\Delta_L}$.
It turned out that the typical characteristic length $\xi_{\Delta_L}$ is about $\sim$ 0.4 nm.
For comparison, we also estimated $\xi_{\Delta_L}$ from our previous results concerning $\Delta$ maps of Bi2212 and Bi2201 (Bi$_{2+x}$Sr$_{2-x}$CuO$_{6+\delta}$) cuprate superconductors\cite{sugimotoPRB2006}.
All estimations are displayed on the $T_c$-$\xi_{\Delta_L}$ plane in Fig. 6(c). 
It is clear from this comparison that the characteristic length $\xi_{\Delta_L}$ of F0256 is much shorter than that of Bi-based cuprate superconductors, i.e. by the factor of 1/2 to 1/3.
Such a short characteristic length seems unexpected, because the spatial-inhomogeneity features of the gap $\Delta_L$ distribution in F0256 turned out to be correlated with the spot structures rather than the superconducting coherence length $\xi$ (several nano-meters).

As was indicated above, the spot structures in F0256 were due to the non-replaced apical oxygen, as comes about from the comparison between the spot number and the O/F ratio.
Therefore, these results indicate that the inhomogeneity should not be associated with superconductivity but is due to some other reasons, such as disorder of BaO/F layers.
Recently, Zeljkovic $et$ $al$ showed that the disordered excess oxygens outside the CuO$_2$ planes behave as the scattering centers and stabilize the pesudogap in Bi2212 \cite{zeljkovic}.
From this, quite reasonable, point of view, one should consider the non-superconducting gap $\Delta_L$ (pseudogap or CDW gap) as strongly affected by the non-replaced oxygen driving the disorder outside of the CuO$_2$ plane, so that the quantity $\Delta_L$ correlates spatially with the spot structures.
Generally speaking, STM/STS measurements probe the surface properties, i.e. mainly those of the outer plane in the studied case.
In the F0256 compound under investigation, the disorder is a result of the replacements of apical O/F in the charge reservoir layer of BaO/F, as can be seen from the appearance of spot structures in STM topographies.
Therefore, outer planes are affected by the disorder of the charge reservoir layer. 
Hence, the STS results are strongly influenced by the properties of the outer plane. 
That is why the order parameter inhomogeneity, i.e the existence of the $\Delta_L$ dispersion ($\sigma \Delta_L $) becomes strongly enhanced.
One can make a conclusion that the length scale of the electronic inhomogeneity found in cuprate superconductors, including both the Bi-based and the apical Fluorine ones, originates from the atomic disorder distributions of the type observed here.

We note here that the magnitudes of $\Delta_L$ gaps are relatively large as compared with those of Bi-based cuprates.
For example, the averaged value of the large gap in F0256 is $\bar{\Delta_L} \sim 79$ meV, while that in the typical optimally doped Bi-based oxide is about $\bar{\Delta} \sim 40$ meV and that in the underdoped one is about $\bar{\Delta} \sim 60$ meV \cite{fujita2012}.
The multi-layered crystal structure of F0256 can be one of the possible reasons of such a discrepancy (see Fig. 1(a)).
Indeed, the studied F0256 superconductor has 6 CuO$_2$ planes in the unit cell.
From the crystal structure of Fig. 1(a), one sees that there are three types of the inequivalent CuO$_2$ planes, the topmost outer plane nearest to the BaO/F layer (OP), the next inner plane (IP1), and the most inner plane (IP2).
According to the NMR studies of multi-layered cuprates, each CuO$_2$ plane possesses a different carrier concentration determined by the distance from the charge reservoir layer of BaO/F, i. e. the inner CuO$_2$ planes have lower carrier density than that of outer CuO$_2$ planes \cite{shimizuNMR_PRL2007,mukuda12:011008}.
In such multi-layered systems, the carrier concentrations per CuO$_2$ plane are reduced.
The gap $\Delta_L$ magnitude, in general, becomes larger with decreasing carrier concentrations in cuprate superconductors, i.e. in underdoped compositions when the samples gradually approach the insulating state.
In fact, according to the recent STS experiments on strongly underdoped Bi2212\cite{hamidian16:Nphys3519}, the averaged $dI/dV$ curve is very similar to ours, namely, demonstrating the existence of a very large gap peak $\Delta_1 \sim$ 100 meV together with the small gap (kink) signal of $\Delta_0$ of $\sim$ 30 meV.

Let us further discuss the existence of multiple gaps in F0256 superconductor.
In the histograms of Fig. 4, only two-peak distributions can be found rather than triple-peak ones, although the F0256 compound possesses 3 types of nonequivalent CuO$_2$ planes.
Similar multiple-gap features were reported for a number of other cuprate superconductors\cite{boyer2201,miyakawa05:225}.
For instance, in the case of Pb-doped Bi2201 oxide, the two-gap patterns were reported, the smaller and larger gap being about 6.7 $\pm$ 1.6 meV and 16 $\pm$ 8 meV, respectively\cite{boyer2201}.
The ingenious trick made it possible to elucidate that the larger gap (the pseudogap) is really inhomogeneous, whereas the inhomogeneity of the smaller (superconducting) gap is rather weak\cite{boyer2201}
Our results also demonstrate that the distribution of the quantity $\Delta_{S-kink}$ is rather homogeneous, while the pseudogap $\Delta_L$ is quite inhomogeneous, indicating that the superconductivity is quite spatially homogeneous against the inhomogeneous pseudogap background.
It should be noted that the previous point-contact spectroscopy revealed the multiple- (triple- or more) gap structure of the electron spectrum, including the superconducting and pseudogap features\cite{miyakawaPCF0234}.
On the contrary, our results show only the double-peak structures.
This discrepancy reflects the difference in the experimental configurations.
Indeed, the STM surface-sensitive measurements detect mainly one or several topmost surfaces.
In particular, our STS experiment seems to probe only the outer CuO$_2$ plane of the exposed surface.
On the other hand, point-contact studies detected all CuO$_2$ planes in the unit cell, including inner and outer planes.
Therefore, their $dI/dV(V)$ curves revealed three or more kinds of gaps.
Previous ARPES studies of the F0234 samples demonstrated the double Fermi surfaces, which was interpreted as the self-doping, the bilayer splitting being associated with the two different types of CuO$_2$ planes, inner and outer ones \cite{chenARPES}.
However, our STS investigations of the F0256 oxide showed no evidence of such splitting for energies in the vicinity of $\Delta_S$.
Nevertheless, the energy-gap edge of $\sim$ 85 meV found in the ARPES measurements\cite{chenARPES} is consistent with our STS large-gap data $\bar{\Delta_L} \sim 79$ meV at 4.9 K.

\section{Summary}
The STM/STS measurements of the multi layered cuprate superconductor Ba$_2$Ca$_5$Cu$_6$O$_{12}$(O$_{1-x}$F$_x$)$_2$ (F0256) were carried out.
The STM topography revealed random spot structures with the spatial scale of $\le 0.5 $ nm. 
The conductance spectra $dI/dV(V)$ were shown to include two kinds of gap features The first, smaller, peculiarity corresponds to the superconducting gap $\Delta_S$, while the other one characterizes the larger gap, i.e. pseudogap $\Delta_L$, indicating the coexistence of superconductivity and pseudogap states down to low temperatures far below $T_c$ within the spatial interval of $\le$ 0.07 nm.
The fitting of the averaged $dI/dV(V)$ curves in the low-voltage, superconducting region, leads to the value $\Delta_{S fit} \sim 14.7$ meV at $T=4.9$ K.
The $T$ dependence of $\Delta_S$ approximately follows the BCS dependence, resulting in the value about 4.9 of the hallmark ratio 2$\Delta_{S}/k_B T_c$, being smaller than those characterizing other cuprates.
The pseudogap $\Delta_L$ observed at low temperatures below $T_c \sim $ 70 K survives above $T_c $ as well. It increases with $T$ from the energy of $\sim 60$ meV to $\sim$ 110 meV, persisting up to 200 K.
The $T$-dependence of the $\Delta$ map below the $T_c$ including small $\Delta_S$ and large $\Delta_L$ was constructed.
From the histogram of the $\Delta$ map, the averaged values at $T=4.9$ K were found to be $\bar \Delta_S = \sim 24$ meV and $\bar \Delta_L = \sim 79$ meV.
The histogram above $T_c$ shows no traces of the double-peak distributions. Only the broad distributions with $\bar \Delta $ $\sim $ 104 meV are observed, clearly indicating that $\Delta_S$ is the superconductive gap.
The spatial distribution of $\Delta_S$ is obtained by means of studying the kink structure of $dI/dV$ curves ($\Delta_{S-kink}$ map).
The $\Delta_S$ shows almost spatially homogeneous distribution, while the pseudogap $\Delta_L$ is quite inhomogeneous, possibly due to the disorder severely affecting the pseudogap-related phenomena.
The $\Delta_L$ maps exhibit the very short characteristic length $\xi_{\Delta_L} \sim 0.4$ nm.
The spatial variation of $\Delta_L$ is correlated with the length scale of the spot structures.
This fact is in favor of the $\Delta_L$ being strongly affected by the disorder of the apical O/F.

\section*{Acknowledgment}
We would thank the Natural Science Center for Basic Research and Development (N-BARD), Hiroshima University for supplying cryogen (liquid helium) and the EPMA element analysis.
This research was supported by JSPS KAKENHI Grant Numbers JP19540370, JP245403770, and the Project N24 of the 2015-2017 Scientific Cooperation Agreement between Poland and Ukraine.


\clearpage


\begin{figure}
\includegraphics[clip,width=170mm]{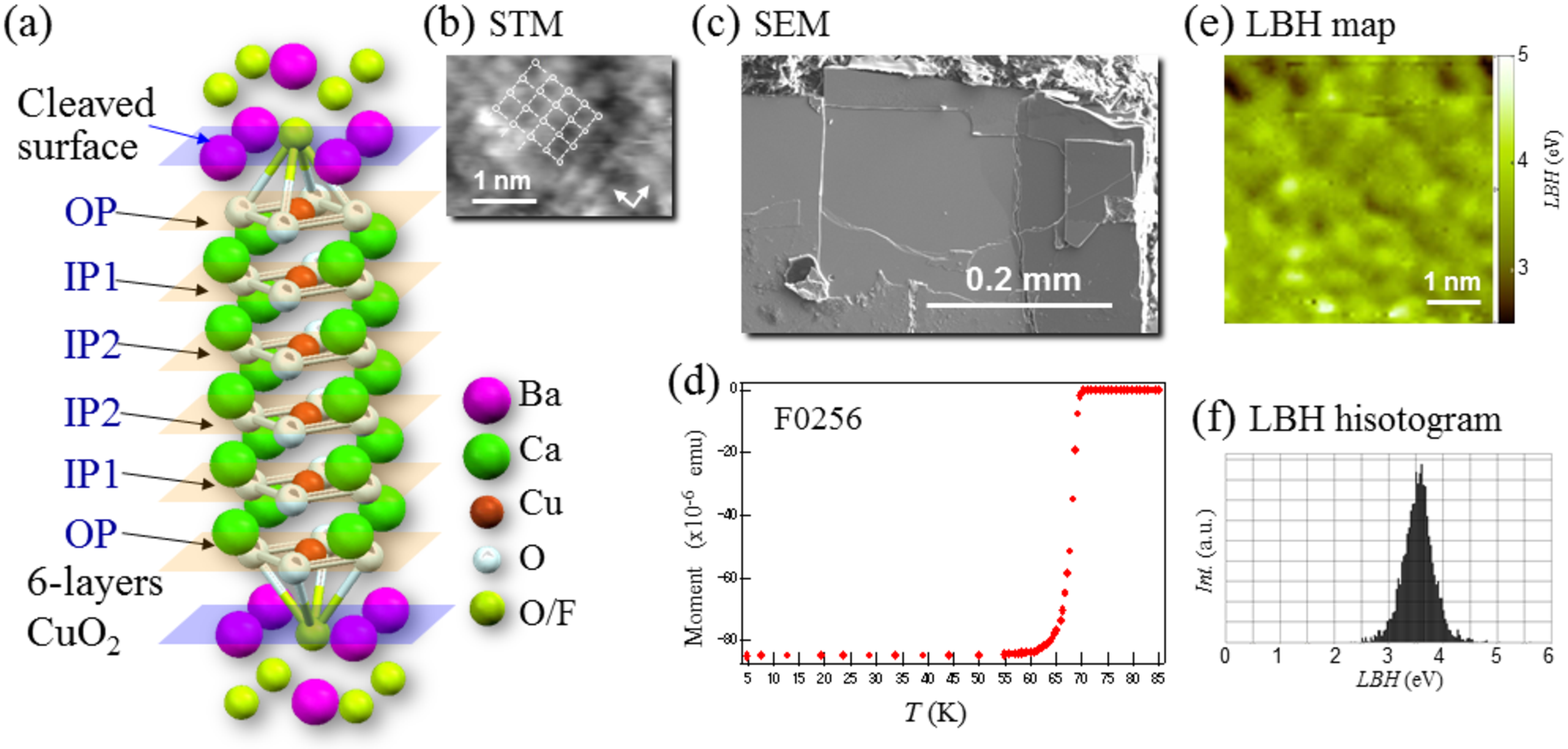}
\caption{(a) The crystal structure of F0256 superconductors. (b) Typical example of the STM topography with the atomic arrangement on a cleaved F0256 single crystal ($V=+0.4$ V, $I=0.2$ nA). (c) SEM micrograph on an F0256 single crystal. (d) Temperature dependence of the magnetic susceptibility. (e) The local barrier height (LBH; work function) map of an F0256 surface (tunneling current $I=0.4$ nA, tip-sample distance range $\Delta z = 0.2$ nm). (f) The histogram of LBH data extracted from the map of Fig. 1(e). }
\label{f1}
\end{figure}

\begin{figure}
\includegraphics[clip,width=170mm]{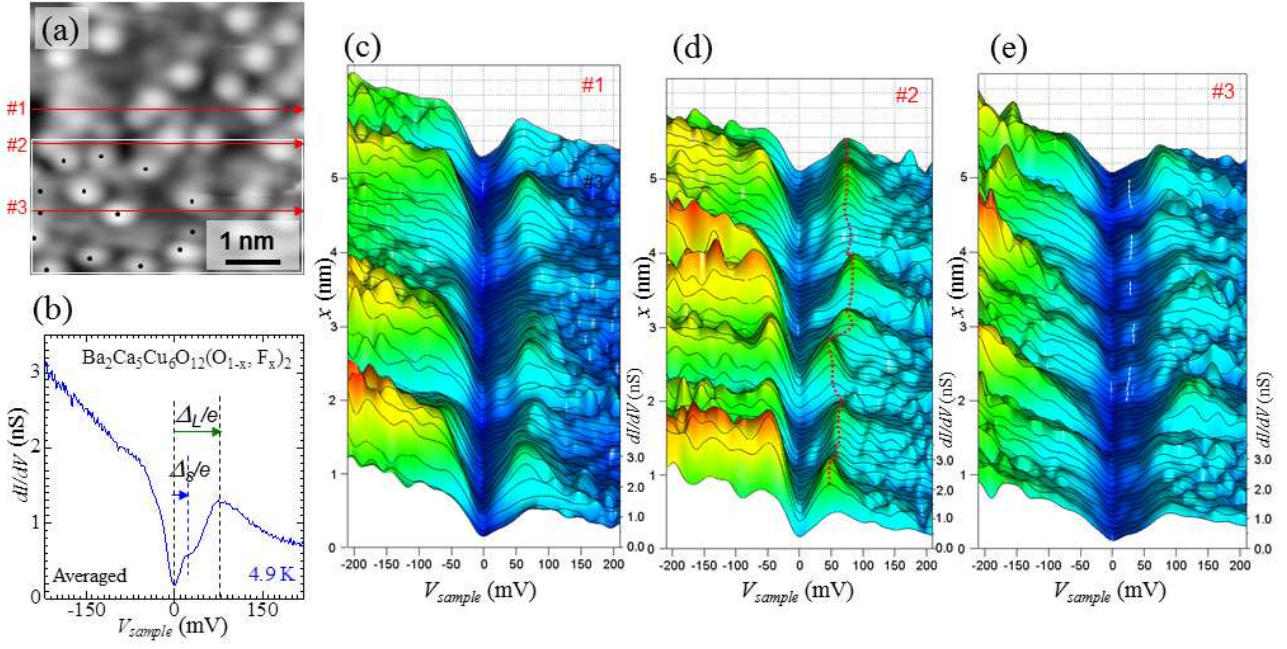}
\caption{(a) The STM topography on F0256 at $T=4.9$ K ($V=+0.3$ V, $I=0.25$ nA). The black dots indicate the positions of the large spots. (b) The averaged $dI/dV(V)$ curve for F0256 at $T=4.9$ K. (c)-(e) The line profiles of the $dI/dV(V)$ spectra at the temperature $T=$4.9 K along arrows $\#$1-$\#$3 in Fig. 2(a).}
\label{f2}
\end{figure}

\begin{figure}
\includegraphics[clip,width=170mm]{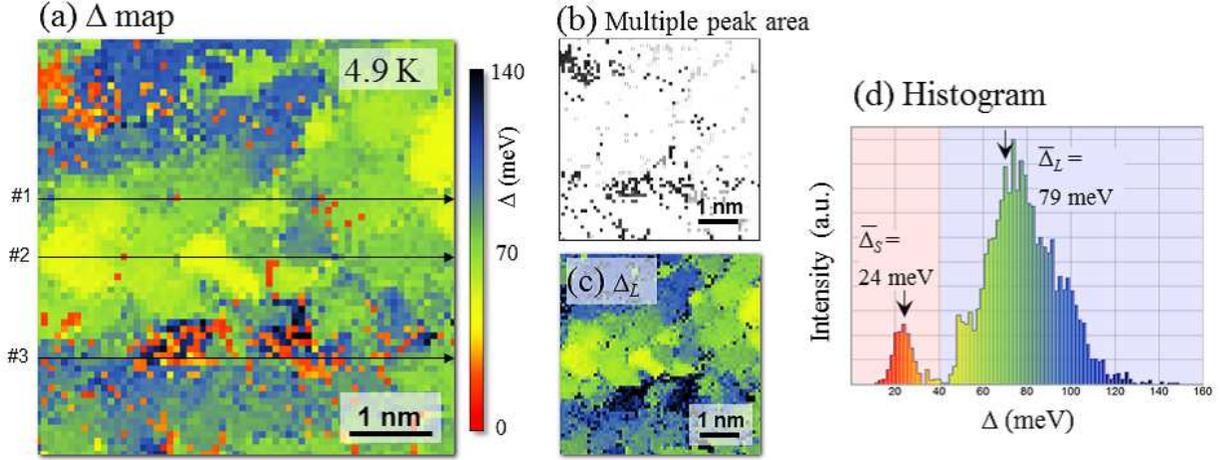}
\caption{(a) The spatial distribution of $\Delta$ ($\Delta$ map) at 4.9K. (b) The multiple gapped area of $\Delta$ map is indicated by gray scale. (c) The large gap ($\Delta_L$) map. The color-scale is the same as that of the Fig. 3(a). The frames of Fig. 3(b) and (c) bound the same area as the $\Delta$ map. The arrows $\#$1 -$\#$3 in Fig. 3(a) indicate the positions of the $dI/dV(V)$ line profiles of Fig. 2(c)$-$(e). (d) The histogram of $\Delta$ extracted from the map of Fig. 3(a).}
\label{f3}
\end{figure}

\begin{figure}
\includegraphics[clip,width=170mm]{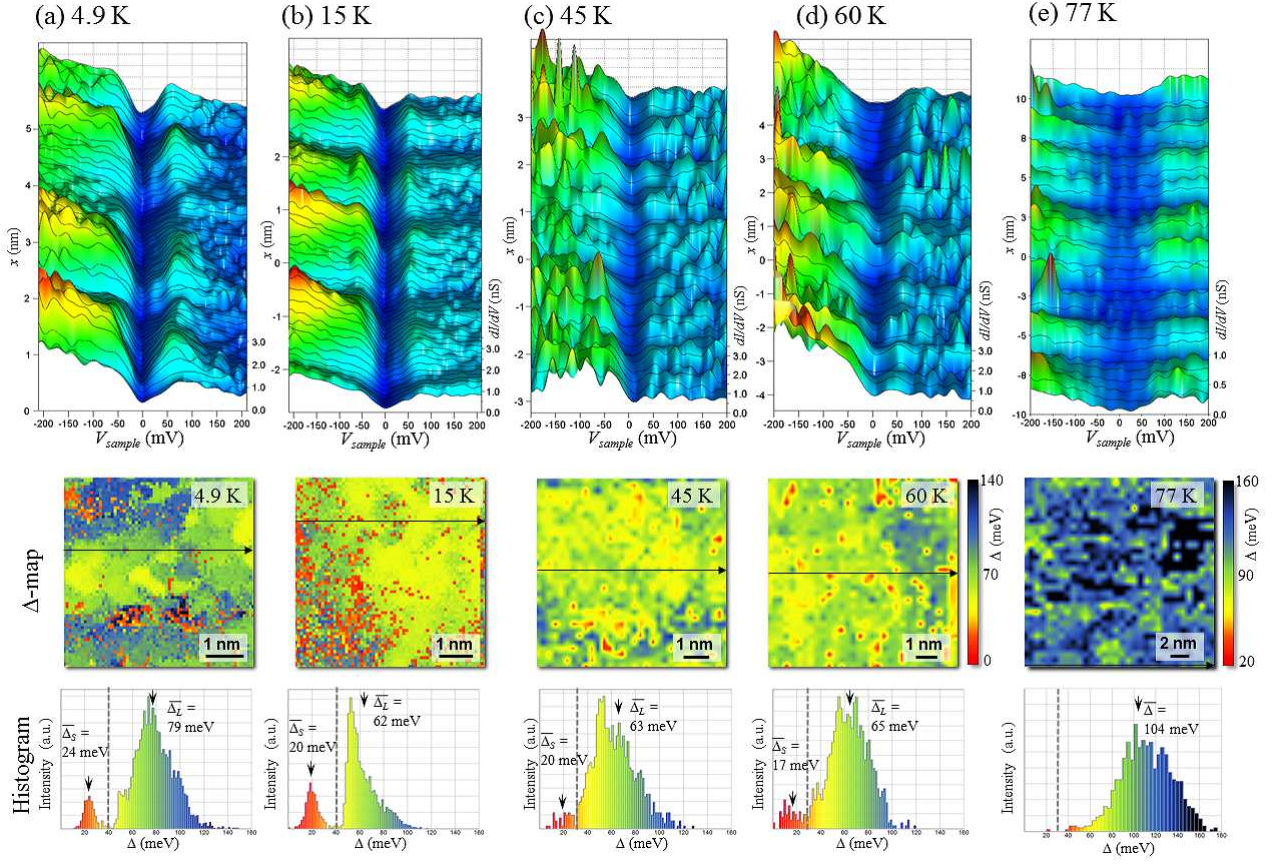}
\caption{The STS spectra at $T$ equals to (a) 4.9, (b) 15, (c) 45, (d) 60, and (e) 77 K, concomitant with the $\Delta$ maps and their histograms. The intervals between measured curves are $\sim 0.07$ nm at $T$=4.9 and 15 K, $\sim 0.2$ nm at $T$=45 and 60 K, and $\sim 0.6$ nm at $T=$77 K. The upper figures show typical examples of the line profiles of the $dI/dV(V)$ spectra. The black arrows in each $\Delta$ map indicate the positions of the $dI/dV(V)$ line profiles for each upper panel.}
\label{f4}
\end{figure}

\begin{figure}
\includegraphics[clip,width=160mm]{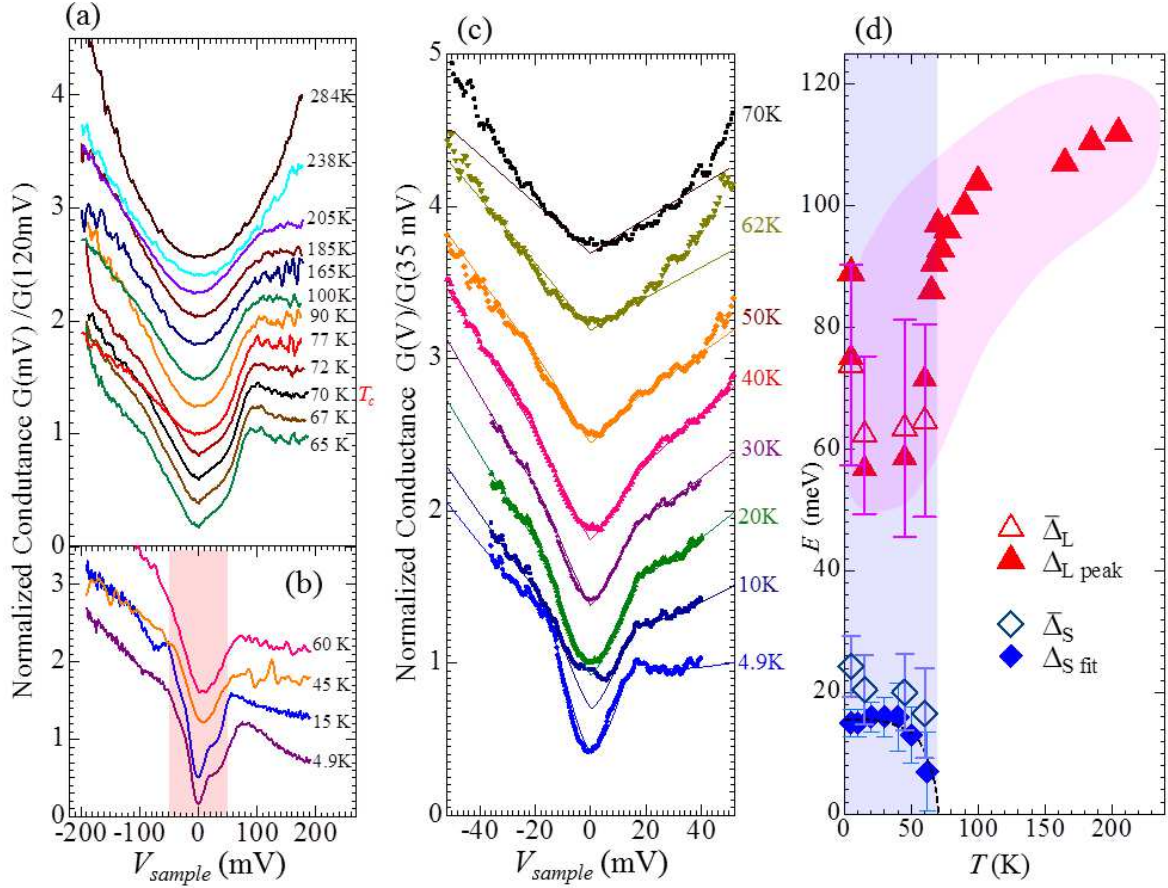}
\caption{The $T$ dependence of the averaged $dI/dV(V)$ curves (a) from 65 K to 284 K, (b) from 4.9 K to 60 K. The curve at $T=$4.9 K in Fig. 5(b) is same as that in Fig. 2(b). (c) The $dI/dV(V)$ curves at low bias voltages from 4.9 K to 70 K. Each $dI/dV(V)$ curve is offset for clarity. The measurements presented in Figs. 5(a) $-$ 5(c) were done using different batches of Pt/Ir tips. (d) The overall $T$ dependence of the gap magnitudes, including both $\Delta_S$ and $\Delta_L$. The dashed line shows the fitting curve $\Delta_{S fit} (T)$ of the $T$ dependence based on the BCS theory.}
\label{f5}
\end{figure}

\begin{figure}
\includegraphics[clip,width=150mm]{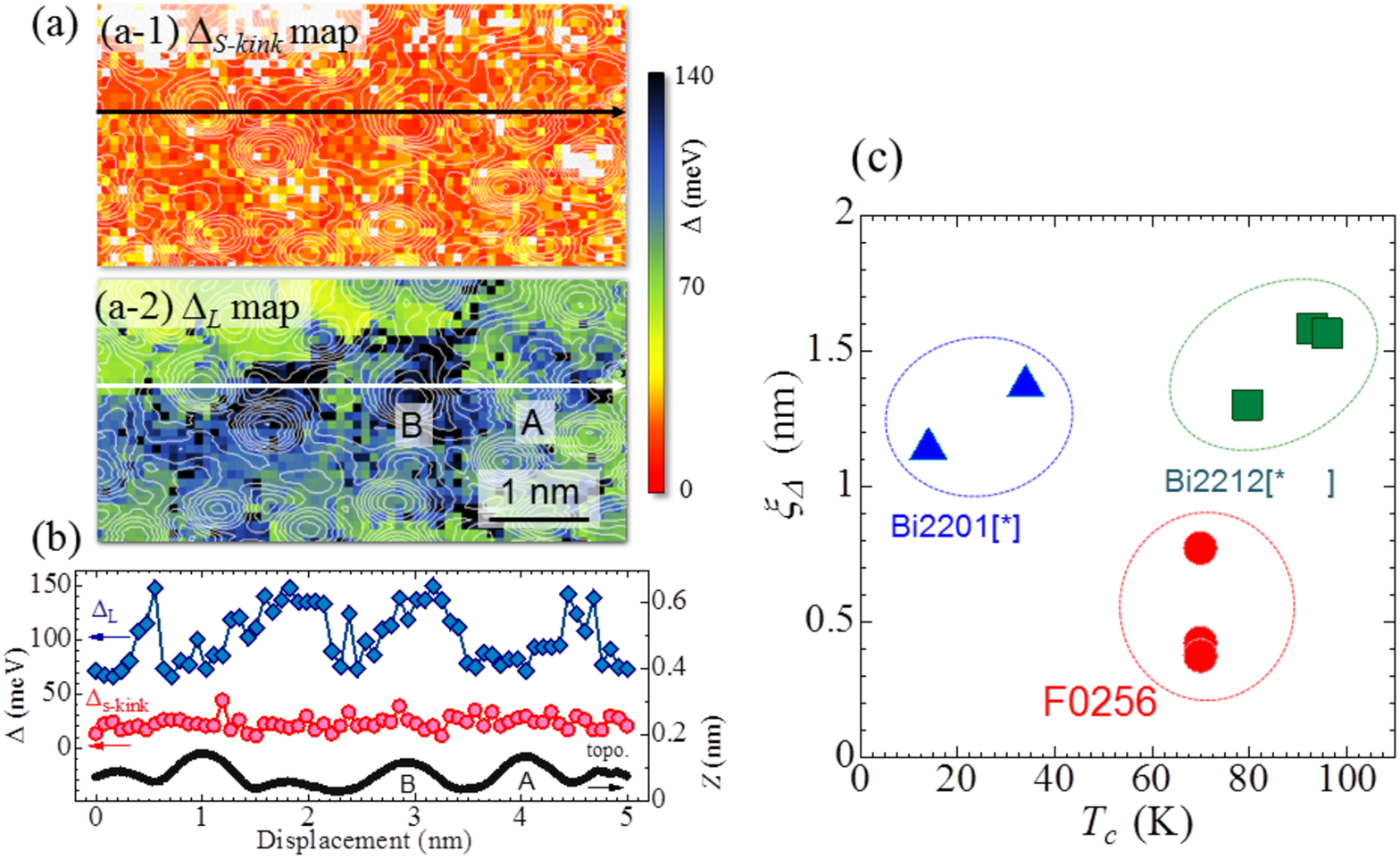}
\caption{The $\Delta_{S-kink}$ map (a-1) and the $\Delta_{L}$ map (a-2) on the squared area of the topography in Fig. 2 (a), with the superposed topography contour plots. The definition of $\Delta_{S-kink}$ value is the inflection point of the kink structures around $V$= +5$\sim$40 mV. (b) The $\Delta_{S-kink}$, $\Delta_L$ and the topography profiles were taken along the arrows of Figs. 6(a-1) and (a-2). (c) The dependence of the characteristic length on $T_c$, $\xi_\Delta$ being defined as the full width at half maximum (FWHM) of the autocorrelation function for the $\Delta$ maps, compared with various cuprate superconductors from our previous report in Ref. [\onlinecite{sugimotoPRB2006}]. }
\label{f6}
\end{figure}


\begin{thebibliography}{100}
\bibitem{klemm12:book} R. A. Klemm, {\it Layered Superconductors}, Volume 1 (University Press, Oxford, 2012).
\bibitem{hirsch15:1} J. E. Hirsch, M. B. Maple, F. Marsiglio, Physica C, {\bf 514}, 1 (2015).
\bibitem{nagamatsu}J. Nagamatsu, N. Nakagawa, T. Muranaka, Y. Zenitani, J. Akimitsu, Nature (London) {\bf 410}, 63 (2001).
\bibitem{kamihara}Y. Kamihara, T. Watanabe, M. Hirano, H. Hosono, J. Am. Chem. Soc. {\bf 130}, 3296 (2008).
\bibitem{hosono15} H. Hosono, K. Kuroki, Physica C, {\bf 514}, 399 (2015).
\bibitem{kubozono16} Y. Kubozono, R. Eguchi, H. Goto, S. Hamao, T. Kambe, T. Terao, S. Nishiyama, L. Zheng, X. Miao, H. Okamoto, J. Phys.: Condens. Matter., {\bf 28}, 334001 (2016).
\bibitem{drozdov15} A. P. Drozdov, M. I. Eremets, I. A. Troyan, V. Ksenofontov, S. I. Shylin, Nature, {\bf 525}, 73 (2015).
\bibitem{bednorz}J. G. Bednorz, K. A. Muller, Z. Phys. {\bf B 64}, 189 (1986). 
\bibitem{keimer15:179} B. Keimer, S. A. Kivelson, M. R. Norman, S. Uchida, J. Zaanen, Nature, {\bf 518}, 179 (2015).
\bibitem{kordyuk15:417} A. A. Kordyuk, Fiz. Nizk. Temp., {\bf 41}, 417 (2015) [Low Temp. Phys., {\bf 41}, 319 (2015)].
\bibitem{mukuda12:011008} H. Mukuda, S. Shimizu, A. Iyo, Y. Kitaoka, J. Phys. Soc. Jpn., {\bf 81}, 011008 (2012).
\bibitem{kresin09:481} V. Z. Kresin, S. A. Wolf, Rev. Mod. Phys., {\bf 81}, 481 (2009).
\bibitem{kamimura12:677} H. Kamimura, H. Ushio, J. Supercond., {\bf 25}, 677 (2012).
\bibitem{hashimoto14:483} M. Hashimoto, I. M. Vishik, R-H. He, T. P. Devereaux, Z-X. Shen, Nat. Phys., {\bf 10}, 483 (2014).
\bibitem{das14:151} T. Das, R. S. Markiewicz, A. Bansil, Adv. Phys., {\bf 63}, 151 (2014).
\bibitem{hufner08:062501} S. H{\"u}fner, M. A. Hossain, A. Damascelli, G. A. Sawatzky, Rep. Prog. Phys., {\bf 71}, 062501 (2008).
\bibitem{yoshida16:014513} T. Yoshida, W. Malaeb, S. Ideta, D. H. Lu, R. G. Moor, Z-X. Shen, M. Okawa, T. Kiss, K. Ishizaka, S. Shin, S. Komiya, Y. Ando, H. Eisaki, S. Uchida, A. Fujimori, Phys. Rev., {\bf B 93}, 014513 (2016).
\bibitem{klemm05:801} R. A. Klemm, Phil. Mag., {\bf 85}, 801 (2005).
\bibitem{kirtley11:436} J. R. Kirtley, C. R. Phys, {\bf 12}, 436 (2011).
\bibitem{lee06:17} P. A. Lee, N. Nagaosa, X-G. Wen, Rev. Mod. Phys., {\bf 78}, 17 (2006).
\bibitem{tajima16:094001} S. Tajima, Rep. Prog. Phys., {\bf 79}, 094001 (2016).
\bibitem{bardeen57:1175} J. Bardeen, L. N. Cooper, J. R. Schrieffer, Phys. Rev., {\bf 108}, 1175 (1957).
\bibitem{won94:1397} H. Won, K. Maki, Phys. Rev., {\bf B 49}, 1397 (1994).
\bibitem{howald}C. Howald, P. Fournier, and A. Kapitulnik, Phys. Rev. {\bf B 64}, 100504R (2001) , C. Howald, H. Eisaki, N. Kaneko, M. Greven, A. Kapitulnik, Phys. Rev. {\bf B 67}, 014533 (2003).
\bibitem{kmlang}K. M. Lang, V. Madhavan, J. E. Hoffman, E. W. Hudson, H. Eisaki, S. Uchida, and J. C. Davis, Nature (London) {\bf 415}, 412 (2002).
\bibitem{matsuda1}A. Matsuda, T. Fujii, and T. Watanabe, Physica {\bf C 388-389}, 207 (2003).
\bibitem{mcelroy}K. McElroy, J. Lee, J.A. Slezak, D. Lee, H. Eisaki, S. Uchida, J.C. Davis, Science {\bf 309}, 1048 (2005).
\bibitem{sugimotoPRB2006}A. Sugimoto, S. Kashiwaya, H. Eisaki, H. Kashiwaya, H. Tsuchiura, Y. Tanaka, K. Fujita, S. Uchida, Phys. Rev. {\bf B 74}, 094503 (2006).
\bibitem{machida}T. Machida, Y. Kamijo, K. Harada, T. Noguchi, R. Saito, T. Kato, and H. Sakata: J. Phys. Soc. Jpn. {\bf 75}, 083708 (2006).
\bibitem{fischer07:353} \O. Fischer, M. Kugler, I. Maggio-Aprile, C. Berthod, Rev. Mod. Phys., {\bf 79}, 353 (2007).
\bibitem{boyer2201}M. C. Boyer, W. D. Wise, K. Chatterjee, M. Yi, T. Kondo, T. Takeuchi, H. Ikuta, E. W. Hudson, Nature Phys {\bf 3}, 802 (2007).
\bibitem{zeljkovic}I. Zeljkovic, Z. Xu, J. Wen, G. Gu, R.S. Markiewicz, J.E. Hoffman, Science {\bf 337}, 320 (2012).
\bibitem{gabovich16:1103} A. M. Gabovich, A. I. Voitenko, Fiz. Nizk. Temp., {\bf 42}, 1103 (2016) [Low Temp. Phys. {\bf 42}, 863 (2016)].
\bibitem{phillips03:2111} J. C. Phillips, A. Saxena, A. R. Bishop, Rep. Prog. Phys., {\bf 66}, 2111 (2003).
\bibitem{damascelli03:473} A. Damascelli, Z. Hussain, Z-X. Shen, Rev. Mod. Phys., {\bf 75}, 473 (2003).
\bibitem{vojta09:699} M. Vojta, Adv. Phys., {\bf 58}, 699 (2009).
\bibitem{armitage10:2421} N. P. Armitage, P. Fournier, R. L. Greene, Rev. Mod. Phys., {\bf 82}, 2421 (2010).
\bibitem{shirage} P. M. Shirage, D. D. Shivagan, Y. Tanaka, Y. Kodama, H. Kito, A. Iyo, Appl. Phys. Lett. {\bf 92}, 222501 (2008).
\bibitem{iyo2001} A. Iyo, Y. Aizawa, K Y. Tanaka, M. Tokumoto, K. Tokiwa, T. Watanabe, H. Ihara, Physica {\bf C 357-360}, 324 (2001).
\bibitem{iyo2003} A. Iyo, M. Hirai, K. Tokiwa, T. Watanabe, Y. Tanaka, Physica {\bf C 392-396}, 140 (2003).
\bibitem{gabovich78:1115} A. M. Gabovich, D. P. Moiseev, Fiz. Nizk. Temp., {\bf 4}, 1115 (1978).
\bibitem{kresin90:1203} V. Z. Kresin, H. Morawitz, Solid State Commun., {\bf 74}, 1203 (1990).
\bibitem{shimizuNMR_PRL2007} S. Shimizu, H. Mukuda, Y. Kitaoka, A. Iyo, Y. Tanaka, Y. Kodama, K. Tokiwa, T. Watanabe, Phys. Rev. Lett. {\bf 98}, 257002 (2007).
\bibitem{miyakawaPCF0234} N. Miyakawa, H. Haya, A. Iyo, Y. Tanaka, K. Tokiwa, T. Watanabe, T. Kaneko, Int. J. Mod. Phys. {\bf B 21}, 3233 (2007).
\bibitem{chenARPES}Y. Chen, A. Iyo, W. Yang, X. Zhou, D. Lu, H. Eisaki, T.P. Devereaux, Z. Hussain, Z-X. Shen, Phys. Rev. Lett. {\bf 97} 236401 (2006).
\bibitem{sugimotoPhysC_ISS2008} A. Sugimoto, K. Shohara, T. Ekino, Y. Watanabe, Y. Harada, S. Mikusu, K. Tokiwa, Y. Watanabe, Physica {\bf C 469}, 1020 (2009).
\bibitem{sugimotoPhysC_ISS2010} A. Sugimoto, R. Ukita, T. Ekino, Y. Harada, T. Furukawa, K. Itagaki, K. Tokiwa, Physica {\bf C 471}, 698 (2011).
\bibitem{ekino2007JPCS} T. Ekino, T. Takasaki, R.A. Ribeiro, T. Muranaka, J. Akimitsu, J. Phys.: Conf. Ser. {\bf 61}, 278 (2007).
\bibitem{sugimotoLBH} A. Sugimoto, T. Ekino, H. Eisaki, J. Phys. Soc. Jpn. {\bf 77}, 043705 (2008).
\bibitem{sugimotoPhysC_M2S}A. Sugimoto, K. Shohara, T. Ekino, Y. Watanabe, Y. Harada, S. Mikusu, K. Tokiwa, Y. Watanabe, Physica {\bf C 470}, S160 (2010).
\bibitem{sugimotoPhysProc_ISS2013}A. Sugimoto, T. Ekino, K. Tanaka, K. Mineta, K. Tanabe, K. Tokiwa, Phys. Proc. {\bf 58}, 71 (2014).
\bibitem{deroo02:PRL097002} D. J. Derro, E. W. Hudson, K. M. Lang, S. H. Pan, J. C. Davis, J. T. Markert, A. L. de Lozanne, Phys. Rev. Lett.{\bf 88}, 097002-1 (2002).
\bibitem{lv15:PRL237002} Y.-F. Lv, W.-L. Wang, J.-P. Peng, H. Ding, Y. Wang, L. Wang, K. He, S.-H. Ji, R. Zhong, J. Schneeloch, G.-D. Gu, C.-L. Song, X.-C. Ma, Q.-K. Xue, Phys. Rev. Lett. {\bf 115}, 237002 (2015).
\bibitem{hamidian16:Nphys3519}M. H. Hamidian, S. D. Edkins, S. H. Joo, A. Kostin, H. Eisaki, S. Uchida, M. J. Lawler, E.-A. Kim, A. P. Mackenzie, K. Fujita, J. Lee, J.C. S. Davis,, Nat. Phys. {\bf 12}, 150 (2016).
\bibitem{gabovich10:681070} A. M. Gabovich, A. I. Voitenko, T. Ekino, Mai Suan Li, H. Szymczak, M. Pekala, Adv. Cond. Mater. Phys., {\bf 2010}, 681070 (2010).
\bibitem{campi15:359} G. Campi, A. Bianconi, N. Poccia, G. Bianconi, L. Barba, G. Arrighetti, D. Innocenti, J. Karpinski, N. D. Zhigadlo, S. M. Kazakov, M. Burghammer, M. v. Zimmermann, M. Sprung, A. Ricci, Nature, {\bf 525}, 359 (2015).
\bibitem{ekino16:445701} T. Ekino, A. M. Gabovich, Mai Suan Li, H. Szymczak, A. I. Voitenko, Journ. Phys.: Condens. Mater., {\bf 28}, 445701 (2016).
\bibitem{yazdani} M. Vershinin, S. Misra, S. Ono, Y. Abe, Y. Ando, A. Yazdani, Science {\bf 303}, 1995 (2004).
\bibitem{ekinoBi}T. Ekino, Y. Sezaki, H. Fujii, Phys. Rev. {\bf B 60} 6916 (1999), T. Ekino, S. Hashimoto, T. Takasaki, H. Fujii, Phys. Rev. {\bf B 64}, 092510 (2001).
\bibitem{matsuda2} A. Matsuda, S. Sugita, T. Watanabe, Phys. Rev. {\bf B 60}, 1377 (1999).
\bibitem{kurosawa10:094519} T. Kurosawa, T. Yoneyama, Y. Takano, M. Hagiwara, R. Inoue, N. Hagiwara, K. Kurusu, K. Takeyama, N. Momono, M. Oda, M. Ido, Phys. Rev., {\bf B 81}, 094519 (2010).
\bibitem{dynes78:1509} R. C. Dynes, V. Narayanamurti, J. P. Garno, Phys. Rev. Lett., {\bf 41}, 1509 (1978).
\bibitem{kirtley92:336} J. R. Kirtley, S. Washburn, D. J. Scalapino, Phys. Rev., {\bf B 45}, 336 (1992).
\bibitem{voitenko10:20} A. I. Voitenko, A. M. Gabovich, Phys. Solid State, {\bf 52}, 18 (2010).
\bibitem{Ekino_Bi_PhysRevB1989}T. Ekino, J. Akimitsu, Phys. Rev. {\bf B 40}, 6902 (1989).
\bibitem{sugimotoPRBMNCl} A. Sugimoto, K. Shohara, T. Ekino, Z. Zheng, S. Yamanaka, Phys. Rev. {\bf B 85}, 144517 (2012).
\bibitem{ekinoNCCO}T. Ekino, A. Sugimoto, S. Hino, K. Shohara, A. M. Gabovich, J. Phys.: Conf. Ser. {\bf 150} 052046 (2009).
\bibitem{zimmers}A. Zimmers, Y. Noat, T. Cren, W. Sacks, D. Roditchev, B. Liang, R.L. Greene, Phys. Rev. {\bf B 76}, 132505 (2007).
\bibitem{kato}T. Kato, S. Okitsu, H. Sakata, Phys. Rev. {\bf B 72}, 144518 (2005).
\bibitem{maki}M. Maki, T. Nishizaki, K. Shibata, T. Sasaki, N. Kobayashi, Physica {\bf C 357-360}, 291 (2001).
\bibitem{gruner94:book} G. Gr{\"u}ner, {\it Density Waves in Solids} (Addison-Wesley, Reading, Massachusetts, 1994).
\bibitem{shen07:216404} D. W. Shen, B. P. Xie, J. F. Zhao, L. X. Yang, L. Fang, J. Shi, R. H. He, D. H. Lu, H. H. Wen, D. L. Feng, Phys. Rev. Lett., {\bf 99}, 216404 (2007).
\bibitem{sinchenko99:4624} A. A. Sinchenko, Yu. I. Latyshev, S. G. Zybtsev, I. G. Gorlova, P. Monceau, Phys. Rev., {\bf B 60}, 4624 (1999).
\bibitem{ekino1986} T. Ekino, J. Akimitsu, Jpn. J. Appl. Phys. {\bf26-3}, 625 (1987).
\bibitem{kordyuk09:020504} A. A. Kordyuk, S. V. Borisenko, V. B. Zabolotnyy, R. Schuster, D. S. Inosov, D. V. Evtushinsky, A. I. Plyushchay, R. Follath, A. Varykhalov, L. Patthey, H. Berger, Phys. Rev., {\bf B 79}, 020504 (2009).
\bibitem{kivelson03:1201} S. A. Kivelson, I. P. Bindloss, E. Fradkin, V. Oganesyan, J. M. Tranquada, A. Kapitulnik, C. Howald, Rev. Mod. Phys., {\bf 75}, 1201 (2003).
\bibitem{fradkin15:457} E. Fradkin, S. A. Kivelson, J. M. Tranquada, Rev. Mod. Phys., {\bf 87}, 457 (2015).
\bibitem{borisenko09:166402} S. V. Borisenko, A. A. Kordyuk, V. B. Zabolotnyy, D. S. Inosov, D. Evtushinsky, B. B{\"u}chner, A. N. Yaresko, A. Varykhalov, R. Follath, W. Eberhardt, L. Patthey, H. Berger, Phys. Rev. Lett., {\bf 102}, 166402 (2009).
\bibitem{achkar12:167001} A. J. Achkar, R. Sutarto, X. Mao, F. He, A. Frano, S. Blanco-Canosa, M. Le Tacon, G. Ghiringhelli, L. Braicovich, M. Minola, M. Moretti Sala, C. Mazzoli, R. Liang, D. A. Bonn, W. N. Hardy, B. Keimer, G. A. Sawatzky, D. G. Hawthorn, Phys. Rev. Lett., {\bf 109}, 167001 (2012).
\bibitem{forgan15:10064} E. M. Forgan, E. Blackburn, A. T. Holmes, A. K. R. Briffa, J. Chang, L. Bouchenoire, S. D. Brown, R. Liang, D. Bonn, W. N. Hardy, N. B. Christensen, M. v. Zimmermann, M. H{\"u}cker, S. M. Hayden, Nature Commun., {\bf 5}, 10064 (2015).
\bibitem{rossnagel11:213001} K. Rossnagel, Journ. Phys.: Condens. Mater., {\bf 23}, 213001 (2011).
\bibitem{mesaros16:1266} A. Mesaros, K. Fujita, S. D. Edkins, M. H. Hamidian, H. Eisaki, S.-i. Uchida, J. C. S. Davis, M. J. Lawler, E.-A. Kim, Proc. Natl. Acad. Sci. USA, {\bf 113}, 12661 (2016).
\bibitem{fujita2012}K. Fujita, A.R. Schmidt, E.-A. Kim, M.J. Lawler, D.H. Lee, J.C.Davis, H. Eisaki, S. Uchida, J. Phys. Soc. Jpn., {\bf 81}, 011005 (2012).
\bibitem{miyakawa05:225} N. Miyakawa, K. Tokiwa, S. Mikusu, T. Watanabe, A. Iyo, J. F. Zasadzinski, T. Kaneko, Int. J. Mod. Phys., {\bf B 19}, 225 (2005).
\end{thebibliography}
\end{document}